\documentclass{svjour3}
\smartqed
\usepackage{amsmath}
\usepackage{graphicx}
\usepackage{dcolumn}
\usepackage{bm}
\usepackage{amssymb}

\begin{document}
\title{Scalar filed evolution and Area Spectrum for Lovelock-AdS Black Holes}
\author{C.B Prasobh\and V.C Kuriakose}
\institute{C.B Prasobh\at Department of Physics, Cochin University of Science and Technology, Cochin 682202, India
\email{prasobhcb@outlook.com}
\and V.C Kuriakose\at Department of Physics, Cochin University of Science and Technology, Cochin 682202, India
\email{vck@cusat.ac.in}
}
\date{Received: 25 June 2013 / Accepted: 30 August 2013}
\maketitle

\begin{abstract}
We study the modes of evolution of massless scalar fields in the asymptotically AdS spacetime surrounding 
maximally symmetric black holes of large and intermediate size in the Lovelock model. It is observed that all modes 
are purely damped at higher orders. 
Also, the rate of damping is seen to be independent of order at higher dimensions. 
The asymptotic form of these frequencies for the case of large black holes is found 
analytically. Finally, the area spectrum for such black holes is found from these asymptotic modes.


\PACS{PACS code1 \and PACS code2 \and more}

\end{abstract}

\section{Introduction}\label{intro}

Gauge-gravity dualities like the AdS/CFT correspondence \cite{maldacena} make it possible to study the properties 
of conformal fields in a particular dimension $d$ by studying the evolution of fields in a black hole spacetime that 
is asymptotically AdS in $(d+1)$ dimensions and this has led to considerable interest in the study of asymptotically 
AdS black hole spacetimes. The main difficulty in studying field evolution in such spacetimes is that the stability 
of the spacetime against perturbations is not always guaranteed, unlike in the case of asymptotically flat spacetimes 
in first order theories in four dimensions. The instability of linear perturbations in higher-order and higher dimensional
theories has already been investigated \cite{laflamme,dotti1,dotti2}. The instability also extends to the thermodynamics of 
the black hole when we consider AdS spacetimes, with the 
well-known Hawking-Page phase transitions \cite{hawking_page} signaling a transition between the black hole spacetime 
and thermal AdS spacetime. Recent studies \cite{takahashiflatunstable,takahashithirdorder,takahashieven} on the Lovelock model 
have confirmed the existence of dynamic instability against metric perturbations in asymptotically flat black hole 
spacetimes. Dynamic instability means that the solutions to the equation for the metric perturbation become unstable 
outside the event horizon for large values of their eigenvalue. This instability exists for all types of metric 
perturbations-tensor, vector and scalar. These instabilities occur when the mass of the black hole falls below a 
lower critical bound that depends on the coupling constants, the dimension of the spacetime and the order of the 
theory. \\ 

Quasi normal modes are damped oscillations (having complex frequencies), known to dominate the intermediate stage of the 
evolution of small perturbations of a black hole spacetime and have been studied extensively. Detailed reviews 
and methods of calculation of quasi normal modes are found in numerous papers that include
\cite{berti,nollert,kokkotas,konoplya1,konoplya2,cho}. Quasi normal modes are known to depend only on the parameters of the 
black hole, such as mass, charge and angular momentum, and be completely independent of the type of the agent that 
caused it. These modes are obtained as the solution of the respective field equation, when solved with respect to the 
metric that describes the particular black hole spacetime of interest. In the case of asymptotically flat spacetimes, 
the effective potential that is perceived by the field in the spacetime often resembles a finite potential barrier such 
that the potential vanishes at infinity. This leads to the possibility of obtaining solutions that are purely ingoing at 
the event horizon of the black hole and purely outgoing at infinity, both resembling plane waves when expressed in terms 
of a scaled co-ordinate called the tortoise co-ordinate. These modes may be observed in the future 
with the aid of gravitational detectors. In the case  of asymptotically AdS spacetimes, however, the 
potential grows indefinitely at infinity. Therefore, one usually looks for solutions that vanish at infinity, while the 
boundary condition at the event horizon unchanged \cite{hubeny}. This is motivated by the case of pure AdS spacetime, where the potential 
effectively confines the field as if ``in a box'' and solutions exist only with a discrete spectrum of real frequencies. 
Even though black holes in asymptotically AdS spacetimes are not believed to exist in nature, interest in studying their quasi 
normal modes stems from the above-mentioned AdS/CFT correspondence \cite{maldacena}.
According to the AdS/CFT correspondence, these perturbations correspond to perturbations of the thermal state of the 
strongly coupled conformal field at the boundary of the spacetime and the quasi normal modes correspond to the 
return to thermal equilibrium, so that the quasi normal frequencies give a measure of the time scale for 
the relaxation, which is difficult to compute directly. This provides the motivation to study the quasi 
normal modes of various fields in asymptotically AdS spacetimes. Earlier work the quasi normal modes of 
Schwarzschild-AdS black holes \cite{hubeny} have proved that the modes scale with the temperature of the 
event horizon.\\

Complete quantization of gravity is one of the major goals of modern theoretical physics. Despite decades of research 
by physicists all over the world, it is yet to be achieved with complete success. One often takes clues from the classical 
theory of a system when attempting its quantization and gravity needs to be no different. Quantization of the black hole 
horizon area is expected to be a major feature of any successful quantum theory of gravity. The original ``classical'' 
theory of gravity, namely the General Theory of Relativity (GTR) does provide us with the tools necessary to estimate 
the value of the area quantum. It is known from field theory that the presence of a periodicity in the classical theory 
of a system points to the existence of an adiabatic invariant with a discrete spectrum in the corresponding quantum 
theory. Interestingly, it has been observed in GTR that the numerical value of these frequencies, in the limit 
of ``large'' frequencies, follow a distinct pattern, with the real part approaching a fixed value. These are termed 
asymptotic frequencies. Suggestions have been made that the fixed value of the real part can be viewed as a physically 
relevant periodicity in the (classical) black hole system which would then lead to the existence of certain adiabatic 
invariant quantity, which in turn would possess equally spaced spectrum according to Bohr-Sommerfeld quantization. Once 
we read it together with  Bekenstein's original proposal \cite{bekenstein} that the black hole entropy is an adiabatic 
invariant with a discrete, equally spaced spectrum, we come to the conclusion that the entropy spectrum (and, by 
extension, the area spectrum) can be deduced from the asymptotic value of the quasi normal frequencies. The connection 
between the fixed asymptotic frequencies and the quantized area spectrum was made by Hod \cite{hod}. 
Dreyer \cite{dreyer} recovered Hod's result in the Loop Quantum Gravity. A new interpretation for the quasi
normal modes $\omega = \omega_R + i \omega_I$ of perturbed black holes 
as equivalent to that of  a collection of damped harmonic oscillators with real frequency 
$\omega_0 = \sqrt{\omega_R^2+\omega_I^2}$ was introduced by Maggiore \cite{maggiore} and used in conjunction 
with Hod's method in order to compute the area spectrum of Schwarzschild black holes.\\

As with the case of quasi normal modes, it is the AdS/CFT correspondence that provides the motivation for 
studying the area spectrum of black holes in asymptotically AdS spacetimes. Recent studies 
(\cite{takayanagi1,takayanagi2,takayanagi3,takayanagi4,dey_roy} and references therein) suggest that 
the gravitational dual of the holographic entanglement entropy in quantum field theories is the area of 
minimal-area surfaces in AdS spaces. The entropy can be used to study phase transitions between various states 
of the field. An area-entanglement entropy relation of the form $S_A = \frac{Area(\gamma_A)}{4G_N^{(d-2)}}$ 
has been proposed \cite{takayanagi1,takayanagi5} which is very similar to the familiar area-entropy relation 
in the General Theory of Relativity. Here, $\gamma_A$ is the $d$-dimensional minimal area surface in 
$AdS_{d+2}$ and ${G_N^{(d-2)}}$ is the $(d+2)$-dimensional
gravitational constant of the AdS gravity. 
Although the above relation was originally proposed for AdS spaces, it is equally applicable to any asymptotically 
AdS space, including one containing a black hole. In that case, the minimal surface tends to wrap the horizon and 
thus we can use the area of the event horizon in order to compute the entanglement entropy in the conformal 
field theory.\\

In this work, we numerically compute the quasi normal frequencies for massless scalar field perturbations 
in asymptotically AdS, spherically symmetric spacetimes in Lovelock model using the 
metric derived in \cite{bhscan}. We analytically find out the asymptotic form of the frequencies 
following \cite{musiri} and use it to deduce the area spectrum of large black holes in the model. A brief outline of 
the paper is as follows: in Sect. \ref{theory}, 
we explain the maximally symmetric Lovelock model and the resulting metric for the spacetime 
as given in \cite{bhscan}. Details of the Horowitz-Hubeny numerical method of computing the 
quasi normal frequencies in AdS black hole spacetimes to the case of massless
scalar fields in the vicinity of such a black hole in the Lovelock-AdS model are also given in the same section. 
The results of the numerical procedure are presented and analyzed in Sect. \ref{discussion}. In Sect. \ref{asymp}, 
we analytically determine the asymptotic quasi normal frequencies 
of the field for the case of large back holes following \cite{musiri}. The results of that analysis are 
used in order to find the area spectrum of the black hole using the Kunstatter's method \cite{kunstatter} in Sect. \ref{area_spectrum}. 
The results are summarized in Sect. \ref{conclusion}.

\section{The metric and the numerical computation of Quasi normal Frequencies}\label{theory}

The Lovelock model \cite{lovelock} is considered to be the most natural generalization of GTR. 
The Lovelock Lagrangian is a polynomial which consists of dimensionally continued higher order curvature terms. 
The most striking property of this Lagrangian is that it yields field equations that are in second order in 
the metric although the Lagrangian itself may contain higher order terms. Also, the theory is known to give solutions 
that are free of ghosts. The maximum order of 
terms in the action, $k$, is fixed by the number of dimensions of the spacetime $d$ in Lovelock model according to the 
relation $k=[\frac{d-1}{2}]$ where $[x]$ denotes the integer part of $x$. The action is written in terms of the 
Riemann curvature $R^{ab}=d\omega ^{ab}+\omega _{c}^{a}\omega ^{cb}$ and the vielbein $e^{a}$ as

\begin{equation}\label{llaction}
I_{G}=\kappa \int \sum_{p=0}^{k}\alpha _{p}L^{(p)},
\end{equation}
where $\alpha _{p}$ are arbitrary (positive) coupling constants, and $L^{(p)}$, given by

\begin{equation}\label{Lovlag}
L^{(p)}=\epsilon _{a_{1}\cdots a_{d}}R^{a_{1}a_{2}}\!\cdot \!\cdot \!\cdot
\!R^{a_{2p-1}a_{2p}}e^{a_{2p+1}}\!\cdot \!\cdot \!\cdot \!e^{a_{d}},
\end{equation}

is the $p^{\textrm{th}}$ order dimensionally continued term in the Lagrangian. The difficulty with the Lagrangian 
given above, with arbitrary values for $\alpha _{p}$, is that it becomes very difficult (if not impossible) to study 
the evolution of fields, since it is not at all clear whether the operator representing the evolution is Hermitian or 
not. As mentioned in the previous section, this problem, for the case of metric perturbations, has been analyzed 
in \cite{takahashiflatunstable}. Although a general instability depending on the black hole mass 
has been established in that work, the numerical value 
for the critical mass has not been calculated. Also, it is very difficult to predict whether the presence of a cosmological 
constant raises or lowers the critical mass, as long as we consider a model with arbitrary $\alpha _{p}$. Moreover, 
for the same case, the the existence of negative energy solutions with horizons and positive energy solutions with 
naked singularities for (\ref{llaction}) has been pointed out earlier \cite{deser,Wheeler}. These difficulties bring out 
the necessity of selecting suitable values for the coupling coefficients in order to have models that support maximally 
symmetric solutions and external perturbations. Maximally symmetric solutions to Lovelock model have long been 
known \cite{bhscan}, which are derived by requiring that the theories must possess a unique cosmological constant 
(and consequently a unique AdS radius $R$) for all orders. The resulting set of coupling constants are seen to be 
labeled by the order $k$ and the gravitational constant $G_k$. The metric describing the spherically symmetric black 
hole spacetime is derived from the action given in (\ref{llaction}) with the choice

\begin{equation}   \label{alphas}
\alpha _{p}=\left\{
\begin{array}{ll}
\frac{R^{2(p-k)}}{(d-2p)}\left(
\begin{array}{c}
k \\
p
\end{array}
\right)  & ,\text{ }p\leq k \\
0 & ,\text{ }p>k
\end{array}
\right.
\end{equation}

where $1\leq k\leq [\frac{d-1}{2}]$. The resulting field equations are of the form

\begin{eqnarray}
\epsilon _{ba_{1}\cdots a_{d-1}}\bar{R}^{a_{1}a_{2}}\!\cdot \!\cdot \!\cdot
\!\bar{R}^{a_{2k-1}a_{2k}}e^{a_{2k+1}}\!\cdot \!\cdot \!\cdot \!e^{a_{d-1}}
&=&0  \label{ssymmfieldeq} \\
\epsilon _{aba_{3}\cdots a_{d}}\bar{R}^{a_{3}a_{4}}\!\cdot \!\cdot \!\cdot \!%
\bar{R}^{a_{2k-1}a_{2k}}T^{a_{2k+1}}e^{a_{2k+2}}\!\cdot \!\cdot \!\cdot
\!e^{a_{d-1}} &=&0  \label{ssymmtorsioneq}
\end{eqnarray}

Here, $\bar{R}^{ab}:=R^{ab}+\frac{1}{R^{2}}e^{a}e^{b}$.
Such theories are labeled by $k$ and have two fundamental constants, $\kappa $ and $R$, 
related to the gravitational constant $G_{k}$ and the cosmological constant $\Lambda $ respectively through the 
relations

\begin{eqnarray}
\kappa &=&\frac{1}{2(d-2)!\Omega _{d-2}G_{k}},  \label{Kappa} \\
\Lambda &=&-\frac{(d-1)(d-2)}{2R^{2}}  \label{Lambda},
\end{eqnarray}

$\Omega _{d-2}$ being the volume of the $(d-2)$ dimensional spherically symmetric tangent space.
The static and spherically symmetric solutions to (\ref{ssymmfieldeq}), written in Schwarzschild-like coordinates, 
take the form

\begin{equation}\label{metric}
ds^{2}=f(r)dt^{2}+\frac{dr^{2}}{f(r)}+r^{2}d\Omega
_{d-2}^{2},
\end{equation}

where $f(r)$ is given by

\begin{equation}\label{f}
f(r) =1+\frac{r^{2}}{R^{2}}-\sigma \left( \frac{C_{1}}{r^{d-2k-1}}\right) ^{1/k}.
\end{equation}

We take $\sigma =1$. The integration constant $C_{1}$ is written as

\begin{equation}\label{c1}
C_{1}=2G_{k}(M+C_{0}),
\end{equation}

where $M$ stands for the mass of the black hole. The constant $C_{0}$ is chosen so that the horizon shrinks to a
point for $M\rightarrow 0$, as

\begin{equation}\label{c0}
C_{0}=\frac{1}{2G_{k}}\delta _{d-2k,1}.
\end{equation}

It is interesting that the exponent of $\left(\frac{1}{r}\right)$ in (\ref{f}) is proportional to $(d-2k-1)$. Since
 $k=[\frac{d-1}{2}]$, $(d-2k-1)=0$ in odd dimensions and $(d-2k-1)=1$ in even dimensions. Thus the solution 
in even dimensions resemble the Schwarzschild-AdS solution. The $(d-2k-1)=0$ cases correspond to Chern-Simmons 
theories which have a vacuum that is different from AdS \cite{bhscan}. Their quasi normal modes, mass and area spectra
have already been computed \cite{gonzalez}. What makes it interesting is the fact that recent 
studies \cite{takahashiflatunstable} on the stability of metric perturbations in Lovelock model also point out that 
it is possible to predict the (in)stability of the perturbations only in even dimensions. The present work is limited 
to the cases where $d-2k-1\neq 0$. Then we have $C_0=0$ and $C_1=2G_kM$. Consider the scalar field $\Phi(r,t,x_i)$ 
that obeys the Klein-Gordon equation given by

\begin{equation}\label{kgeq}
\frac{1}{\sqrt{g}} \partial_A \sqrt{g}g^{AB}\partial _B \Phi = 0,
\end{equation}

$x_i$ being the co-ordinates in the spherically symmetric tangent space and $g_{AB}$ being components of the 
metric tensor. We impose the boundary conditions of ingoing plane wave solution
at the event horizon and vanishing field at the boundary. The boundary condition at the horizon suggests the 
ansatz $\Phi = e^{-i\omega (t+r_*)}$ where $r_*$ is the tortoise coordinate defined by

\begin{equation}\label{tort}
dr_*=\frac{dr}{f(r)}.
\end{equation}

In the $(v=t+r_*,r)$ system, the metric reads

\begin{equation}\label{infalling_metric}
 ds^2 = - f(r)dv^2 + 2dvdr+r^2d\Omega_{d-2}^2,
\end{equation}

and we take the ansatz

\begin{equation}\label{phi_infall_ansatz}
 \Phi(v,r,x_i) =  r^{\frac{2-d}{2}}\psi(r)Y(x_i) e^{-i\omega v},
\end{equation}

so that (\ref{kgeq}) becomes

\begin{equation}\label{hubeny_eq_r}
 f(r)\frac{d^2}{dr^2} \psi(r)
    + [f'(r) - 2 i\omega]  \frac{d}{dr} \psi(r) 
    - V(r)  \psi(r) = 0,
\end{equation}

with the effective potential

\begin{equation}\label{hubeny_pot_r}
 V(r) = \frac{(d-2)(d-4)}{4r^2}f(r) + \frac{d-2}{2r} f'(r) + \frac{l(l+d-3)}{r^2}.
\end{equation}

Here, $l$ represents the eigenvalue of the operator on the LHS of (\ref{kgeq}) acting on the functions $Y(x_i)$ in 
the spherically symmetric tangent space. In order to numerically calculate the quasi normal frequencies for 
(\ref{hubeny_eq_r}), we expand the field $\Phi$ as a power series about the horizon and impose the vanishing 
boundary condition at infinity. We change the variable from $r$ to $x=\frac{1}{r}$ in order to map the range  
$r_+ < r < \infty$ to a finite range. In terms of $x$, (\ref{hubeny_eq_r}) becomes

\begin{equation}\label{hubeny_eq_x}
 s(x) \, \frac{d^2}{dx^2} \psi(x)
    + \frac{t(x)}{x-x_+}  \, \frac{d}{dx} \psi(x) 
    + \frac{u(x)}{(x-x_+)^2}  \, \psi(x) =0,
\end{equation}

where 

\begin{eqnarray}
 &s(x)=-x^4f(x),~t(x)=-x^4f^\prime(x)-2x^3f(x)-2i\omega x^2,\nonumber\\&u(x)=(x-x_+)V(x)\label{s_t_u_def},
\end{eqnarray}

and

\begin{eqnarray}
 V(x)=\frac{(d - 2)(d - 4)x^2f(x)}{4} - \frac{(d - 2)x^3f^\prime(x)}{2}\nonumber\\ + l(l + d - 3)x^2. \label{hubeny_pot_x}
\end{eqnarray}

In the numerical procedure, we find out the coefficients of the expansion of the functions $s(x),~t(x)$ and $u(x)$ 
as power series in $(x-x_+)$ using a computer with $s_i,~t_i$ and $u_i$ denoting the coefficient for 
the $i^{\textrm{th}}$ term. Then an expansion for $\psi(x)$ of the form

\begin{equation}\label{psi_x_ansatz}
 \psi(x) = \sum_{n=0}^{\infty} a_n  (x - x_+)^n,
\end{equation}

is substituted into (\ref{hubeny_eq_x}) which yields 

\begin{equation}\label{coeft_an}
 a_n = -\frac{1}{P_n} \sum_{k=0}^{n-1} [ k (k -1) s_{n-k} + k t_{n-k} + u_{n-k} ]  a_k,
\end{equation}

where

\begin{equation}\label{coeft_pn}
 P_n = n (n-1) s_0 + n t_0.
\end{equation}

We fix $a_0$ and numerically calculate the coefficients in (\ref{s_t_u_def}) and (\ref{coeft_an}) to different 
orders and compute the value of the filed $\psi(x)$ as given in (\ref{psi_x_ansatz}). Since we wish to impose 
the boundary condition of vanishing field at infinity, we solve the equation $\psi(0)=0$ for $\omega$. It is 
observed, by comparison with the values in \cite{hubeny}, that the quasi normal frequencies are one of the solutions 
of the equation $\psi(0)=0$, solved for $\omega$, after $\psi$ has been computed using (\ref{psi_x_ansatz}) for some 
reasonably high value of $n$, which should be fixed by trial and error. We assume the same to be true in higher orders 
as well and look for the value of the quasi normal modes among the set of discrete values of $\omega$ that are 
obtained. For each value of $\omega$ obtained, we evaluate the absolute value of the LHS of (\ref{hubeny_eq_x}) 
at a point close to the event horizon, since we have assumed plane wave solutions there. We choose that value 
of $\omega$ as the quasi normal frequency for which the absolute value comes closest to zero (the assumption here 
is that (\ref{psi_x_ansatz}) is satisfied exactly only for the quasi normal frequency, which we seek, and not by 
other roots of $\psi=0$.). We increase the number of terms to which $\psi(x)$ and the coefficients are calculated 
until the required precision is attained.

\section{Discussion on Results of the Numerical Calculation}\label{discussion}

\begin{figure}[h]
\includegraphics[scale=0.5]{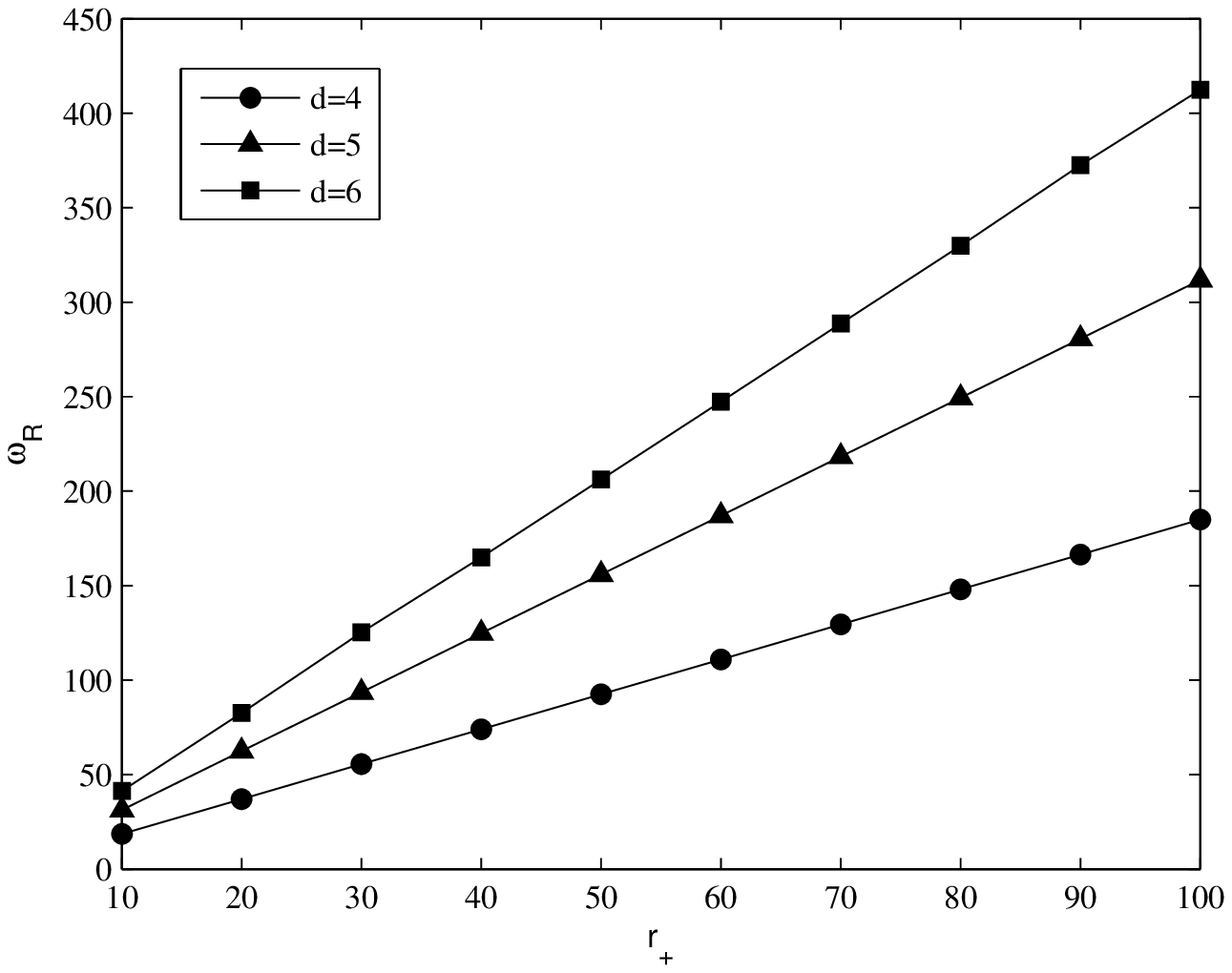}
\caption{\label{fig:orlfo} $\omega_R$ vs $r_+$ plot for large black holes in first order theories}
\end{figure}

\begin{figure}[h]
\includegraphics[scale=0.5]{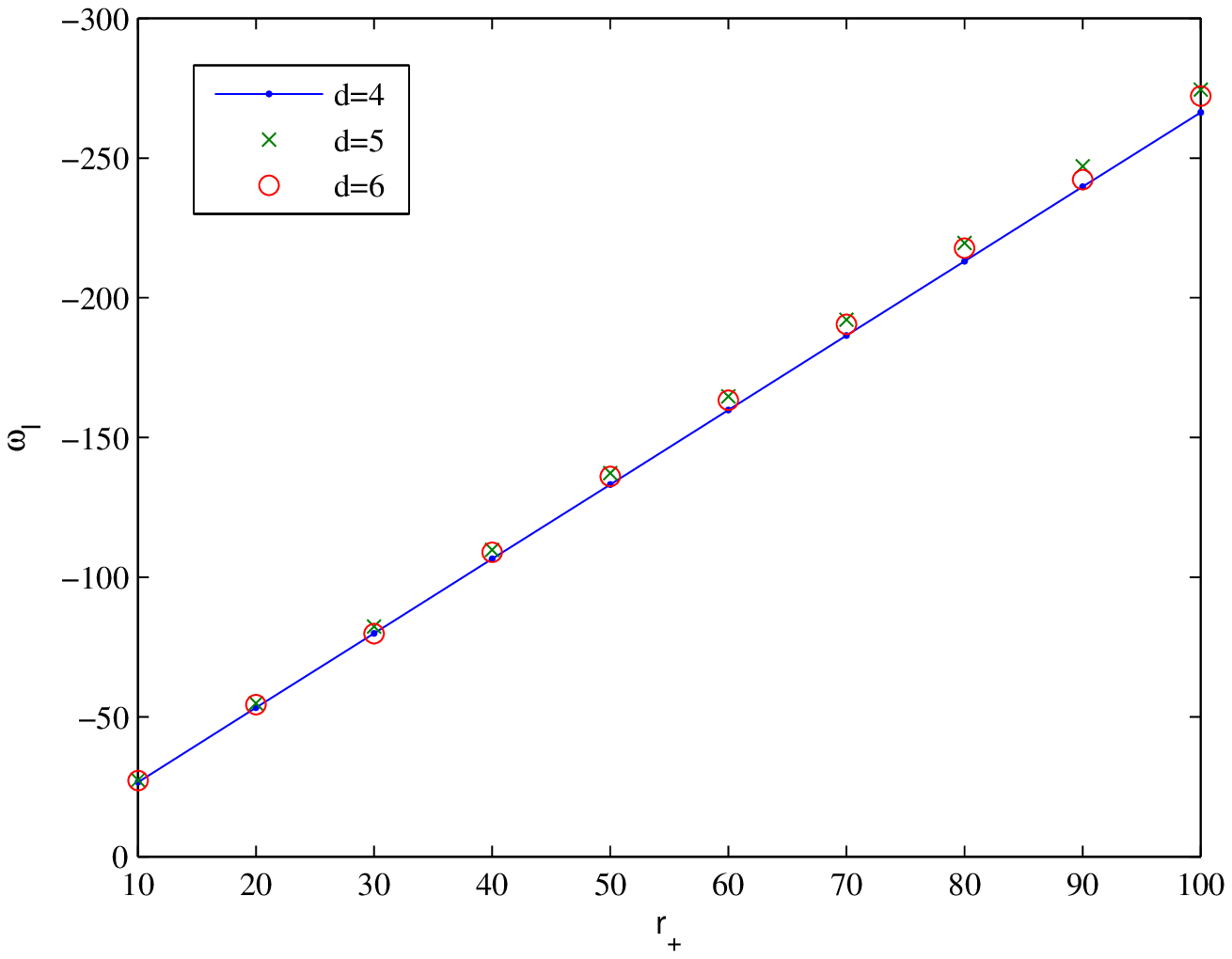}
\caption{\label{fig:oilfo} $\omega_I$ vs $r_+$ plot for large black holes in first order theories}
\end{figure}

\begin{figure}[h]
\includegraphics[scale=0.5]{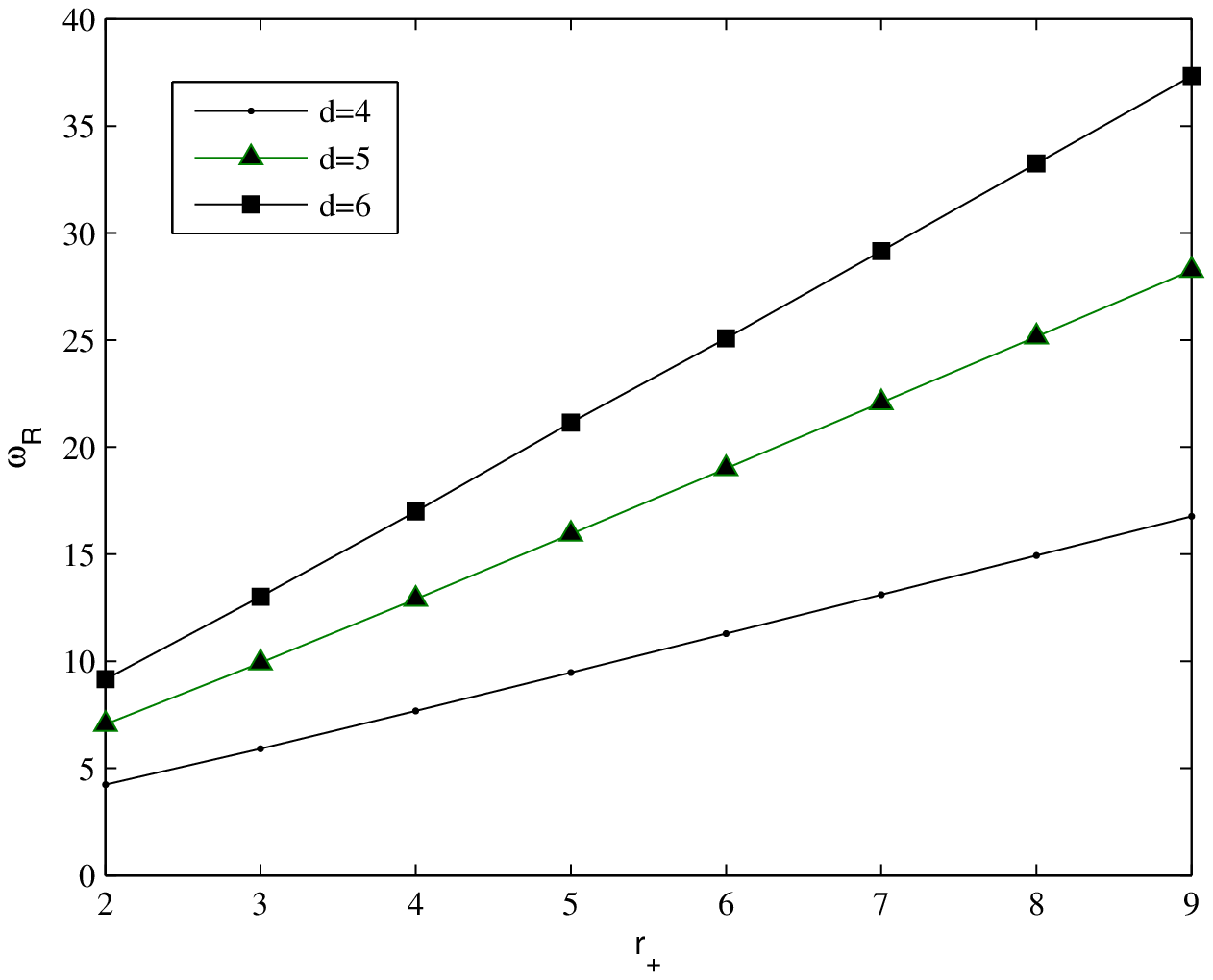}
\caption{\label{fig:orifo} $\omega_R$ vs $r_+$ plot for intermediate black holes in first order theories}
\end{figure}

\begin{figure}[h]
\includegraphics[scale=0.5]{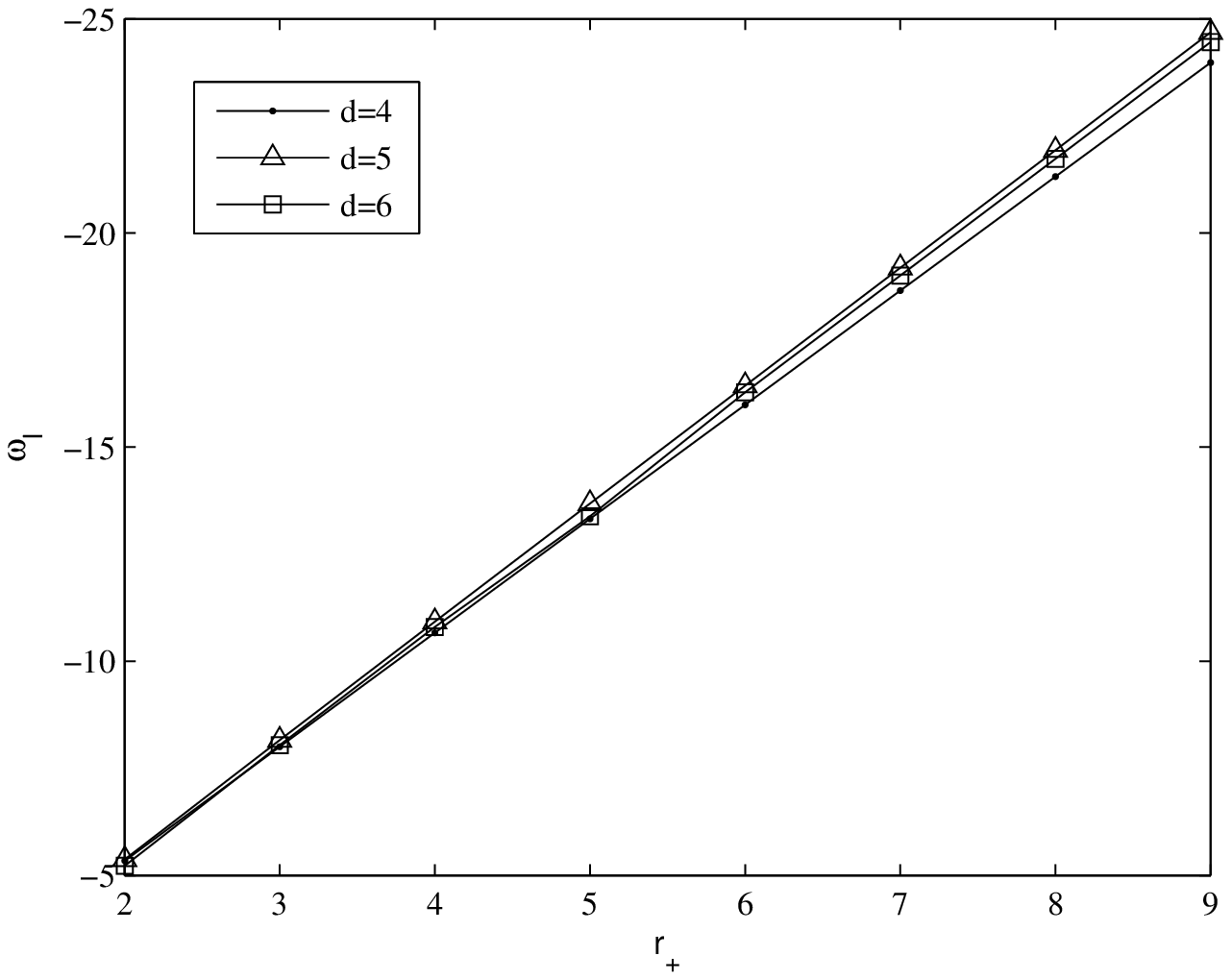}
\caption{\label{fig:oiifo} $\omega_I$ vs $r_+$ plot for intermediate black holes in first order theories}
\end{figure}

\begin{figure}[h]
\includegraphics[scale=0.5]{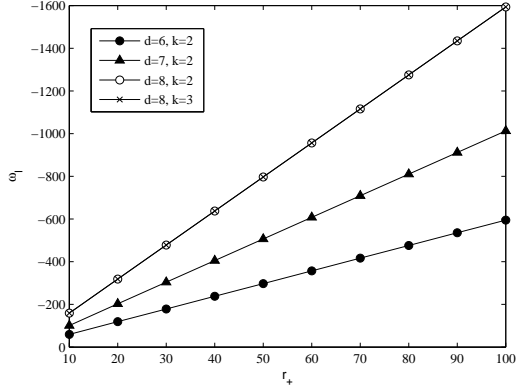}
\caption{\label{fig:oilho} $\omega_I$ vs $r_+$ plot for large black holes in higher order theories}
\end{figure}

\begin{figure}[h]
\includegraphics[scale=0.5]{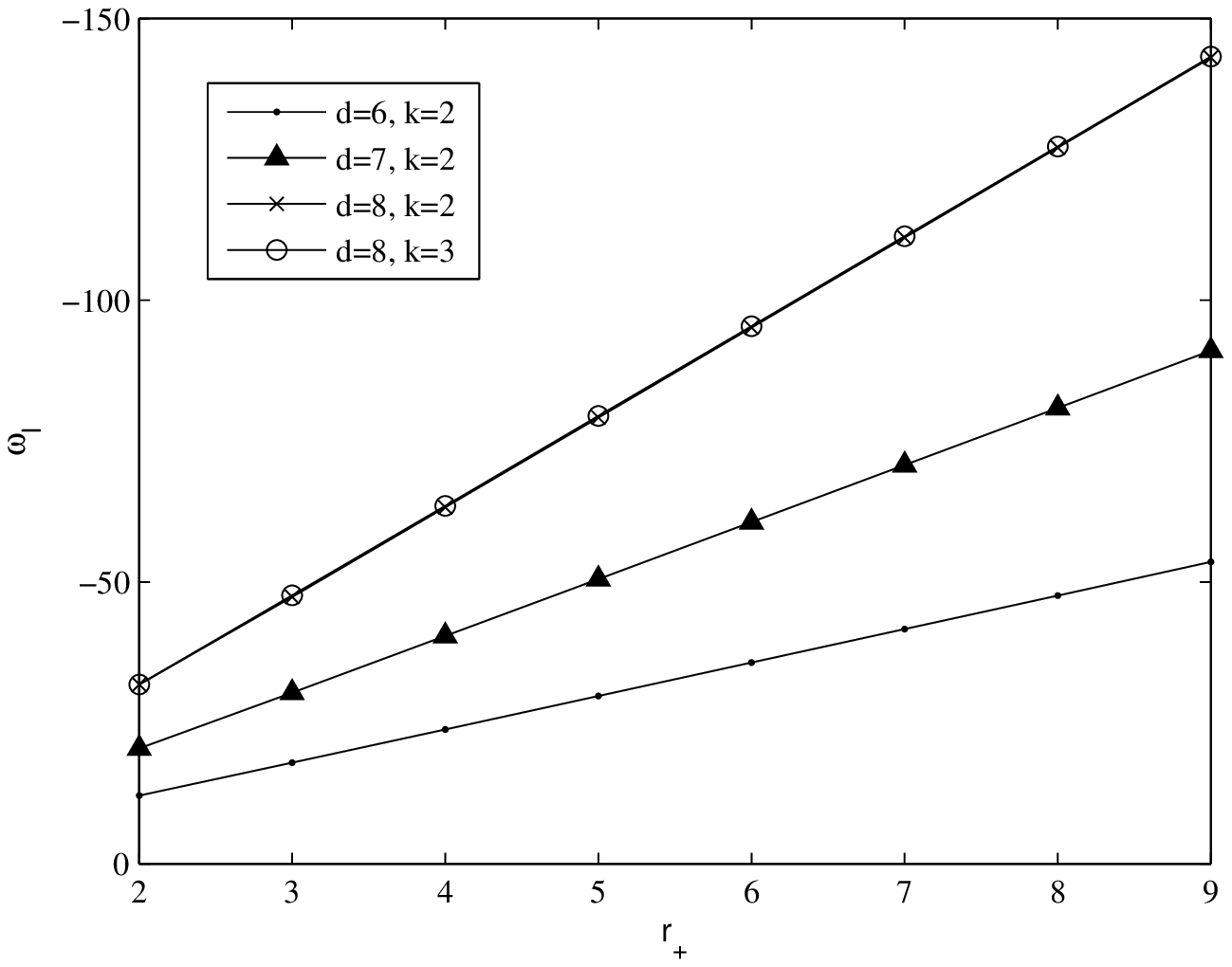}
\caption{\label{fig:oiiho} $\omega_I$ vs $r_+$ plot for intermediate black holes in first order theories}
\end{figure}

We implement the procedure outlined above after fixing the value of the constants $a_0$ and $R$ to $1$.
 We investigate the quasi normal frequencies of large $(r_h \gg R)$ and intermediate $(r_+ \sim R)$ black holes and 
set $\omega = \omega_R - i \omega_I$ as done in \cite{hubeny} since we are interested in damped modes. As mentioned
 before, the analysis is limited to the cases where $d-2k-1 \neq 0$. The results for the lowest $(l=0)$ modes of the
 massless scalar field for first order theories have been summarized in TABLE \ref{tab:table1}. TABLE \ref{tab:table2} 
contains the same for higher orders. Figures \ref{fig:orlfo} to \ref{fig:oiiho} show the results in detail. 
As evident from Figure \ref{fig:orlfo} and Figure \ref{fig:oilfo}, both $\omega_R$ and $\omega_I$ show linear dependence 
on $r_+$ for the case of large black holes in first order 
theories. The linearity seems to be broken when we move to the intermediate-sized black holes, the results for which 
have been plotted in 
Figure \ref{fig:orifo} and Figure \ref{fig:oiifo}. Although the plots for intermediate-sized black holes look linear, 
the $\omega_R-r_+$ dependence for their case rather resembles an $(x,y=x+\frac{1}{x})$ relation. The temperature of the 
event horizon 
for the metric (\ref{metric}), given by

\begin{equation}\label{temperature}
 T=\frac{1}{4 \pi \kappa_B k}\left( \left( d-1 \right)\frac{r_+}{R^2} + \frac{ d-2k-1 }{r_+} \right),
\end{equation}

where $\kappa_B$ denotes the Boltzmann's constant, also depends on $r_+$ in the same manner. 
Since $(x,y=x+\frac{1}{x}) \sim x $ for large $x$, we conclude that both $\omega_R$ and $\omega_I$ scale with the 
temperature 
for large as well as intermediate-sized black holes for first order theories, in agreement with earlier 
works \cite{hubeny}. 
Another observation is that $\omega_I$ seems to be independent of dimension $d$ in first order theories. 
When we consider higher order theories, the numerical results for which have been plotted in Figure \ref{fig:oilho} 
and Figure 
\ref{fig:oiiho}, we observe that all modes are purely damped ones. We have only plotted $\omega_I$ vs $r_+$ in the case 
of higher order theories for this reason. There, we observe that, both for large and intermediate-sized black holes, 
$\omega_I$ is independent of the order of the theory when the dimension $d$ stays the same.

\begin{table*}
\caption{\label{tab:table1}Variation of the frequency of the lowest $(l=0)$ massless ($m=0$) mode for first order theories.}
\begin{tabular}{ccccccccc}

 \\&\multicolumn{2}{c}{$d=4,k=1$}&\multicolumn{2}{c}{$d=5,k=1$}&\multicolumn{2}{c}{$d=6,k=1$}\\\\\hline\\
 $r_+$&$\omega_R$&$\omega_I$&$\omega_R$&$\omega_I$&$\omega_R$&$\omega_I$\\\\ \hline\\
 
 100 & 184.958 & -266.392 & 311.785 & -274.542 & 412.327 & -272.185\\
 
 50 & 92.496 & -133.196 & 155.919 & -137.268 & 206.194 & -136.088\\
 
 10 & 18.608 & -26.642 & 31.351 & -27.432 & 41.434 & -27.187\\
 
 5 & 9.471 & -13.326 & 15.935 & -13.683 & 21.146 & -13.386\\
 
 3 & 5.916 & -8.001 & 9.921 & -8.161 & 13.015 & -8.044\\
 
 2 & 4.235 & -5.340 & 7.062 & -5.376 & 9.164 & -5.223
 
\end{tabular}
\end{table*}

\begin{table*}
\caption{\label{tab:table2}Variation of the frequency of the lowest $(l=0)$ massless ($m=0$) mode in higher orders.}
\begin{tabular}{ccccccccc}
 \\&\multicolumn{2}{c}{$d=6,k=2$}&\multicolumn{2}{c}{$d=7,k=2$}&\multicolumn{2}{c}{$d=8,k=2$}&\multicolumn{2}{c}{$d=8,k=3$}\\\\\hline\\
 $r_+$&$\omega_R$&$\omega_I$&$\omega_R$&$\omega_I$&$\omega_R$&$\omega_I$&$\omega_R$&$\omega_I$\\\\ \hline\\
 
 100 & 0 & -595.081 & 0 & -1013.063 & 0 & -1593.741 & 0 & -1593.762\\
 
 50 & 0 & -357.045 & 0 & -506.450 & 0 & -796.796 & 0 &  -796.839\\
 
 10 & 0 & 26.642 & 0 & -101.185 & 0 & -159.019 & 0 & -159.205\\
 
 5 & 0 & -29.776 & 0 & -50.523 & 0 & -79.193 & 0 & -79.467\\
 
 3 & 0 & -17.938 & 0 & -30.354 & 0 & -47.368 & 0 & -47.623\\
 
 2 & 0 & -12.105 & 0 & -20.475 & 0 & -31.807 & 0 & -31.831

\end{tabular}
\end{table*}

\section{Asymptotic quasi normal modes and area spectrum of large black holes}\label{asymp}

We analytically find the asymptotic form of the quasi normal frequencies in the large black hole limit following the 
method of perturbative expansion of the wave equation in the dimensionless parameter $\omega/T_H$ that has earlier been
 employed in the case of $d$-dimensional SAdS black holes \cite{musiri}. Here, $T_H$ is the Hawking temperature of the 
horizon and $\omega$ is the frequency of the mode. We take the metric to be of the form given in (\ref{metric}). For 
large black holes, the metric gets approximated as,

\begin{equation}\label{metric_approx}
ds^{2}=\hat f(r)dt^{2}+\frac{dr^{2}}{\hat f(r)}+r^2 ds^2 (\mathbb{E}^{d-2}),
\end{equation}

with

\begin{equation}\label{f_approx}
\hat f(r) = \frac{r^2}{R^2} - \left( \frac{2G_kM}{r^{d-2k-1}}\right).
\end{equation}

It is easily seen that the event horizon is given by

\begin{equation}\label{horizon}
 r_h = R \left[ \frac{2G_kM}{R^{d-2k-1}} \right] ^\frac{1}{d-1}.
\end{equation}

In terms of the new metric (\ref{metric_approx}), the Klein-Gordon field equation (\ref{kgeq}) for $m=0$ becomes

\begin{equation}\label{kgeq_approx}
 \frac{1}{r^{d-2}}\partial_r (r^d A(r) \partial_r \Phi) -\frac{R^4}{ r^2 A(r) }\partial_{t}^2\Phi - \frac{R^2}{r^2} \nabla^2 \Phi = 0,
\end{equation}

where 

\begin{equation}\label{A}
 A(r)=1-\left(\frac{r_h}{r}\right)^\frac{d-1}{k}.
\end{equation}

We write the field $\Phi$ as

\begin{equation}\label{Phi_ansatz}
 \Phi(t,r,x_i) = e^{i(\omega t - \vec p\cdot \vec x)} \Psi (r),
\end{equation}

and change the variable from $r$ to 

\begin{equation}\label{y}
 y=\left(\frac{r}{r_h}\right)^\frac{d-1}{2k},
\end{equation}

so that (\ref{kgeq_approx}) becomes

\begin{eqnarray}
y^Q (y^2-1)\left( y^{2k-1}(y^2-1) \Psi' \right)' \nonumber \\
+ \left[\frac{\hat\omega^2}{A^2}\, y^2 - \frac{\hat p^2}{A^2} (y^2-1)\right]\Psi = 0 \label{field_eqn_y},
\end{eqnarray}

where the parameters $\hat\omega$ and $\hat p$ are defined as 

\begin{equation}\label{dimless_param}
 \hat\omega = \frac{\omega R^2}{r_h},~~~~~
\hat p = \frac{|\vec p|R}{r_h},
\end{equation}

and

\begin{equation}\label{Q_and_A}
 Q=\frac{6k-(2k-1)d-1}{d-1},~A=\frac{d-1}{2k}.
\end{equation}

We investigate the behavior of (\ref{field_eqn_y}) near the boundaries $y\to 1$ and $y\to \infty$ and the 
point $y\to -1$ in order to develop an ansatz for $\Psi(y)$. The following solutions are obtained:

\begin{equation}\label{boundary_condition}
\Psi \sim \left\{
\begin{array}{lll}
y^{-2k} & ,\text{ }y \to \infty \\

(y-1)^{\pm i \hat \omega / 2A}  & ,\text{ }y \to 1 \\

(y+1)^{\pm \hat \omega / 2A} & ,\text{}y \to -1
\end{array}
\right.   
\end{equation}

Since we demand ingoing plane wave like solutions at the horizon, we take the form 
$\Psi \sim (y-1)^{-i \hat \omega / 2A}$ near the horizon $(y=1)$.  We isolate the solutions near $y=\pm 1$ and write 

\begin{equation}\label{Psi_ansatz_pm1}
 \Psi (y) = (y-1)^{-i\hat\omega/2A} (y+1)^{\pm\hat\omega/2A} N(y).
\end{equation}

Substituting (\ref{Psi_ansatz_pm1}) in (\ref{field_eqn_y}), we deduce the equation satisfied by $N(y)$ as 

\begin{eqnarray}\label{heun}
y(y^2-1) N'' - \frac{\hat \omega^2 y^2}{A^2(y^2-1)}N +\nonumber\\+ \left\{ \frac{\hat\omega}{A}\left( \mp \frac{i\hat\omega}{2A} \pm k-ik\right) y - (i\pm 1)(2k-1)\frac{\hat\omega}{2A}  \right\} N \nonumber\\ \left\{ \left( 2k+1- \frac{i\mp 1}{A} \hat\omega \right) y^2 - \frac{i \pm 1}{A} \hat\omega y -(2k-1) \right\} N' \nonumber \\
+\frac{1}{y^{Q+2k-2}}\left(\frac{\hat \omega^2 y^2}{A^2(y^2-1)}- \frac{\hat p^2}{A^2}\right)N = 0.
\end{eqnarray}

We consider (\ref{heun}) in the range of large $\hat \omega$ and large $y$, so that $y^2\approx y^2-1$ and the terms 
proportional to $1/(y^{Q+2k-2})$ may be dropped along with the constant terms. Then (\ref{heun}) reduces to

\begin{eqnarray}\label{hypergeom}
(y^2-1) N'' +  \frac{\hat\omega}{A}\left( \mp \frac{i\hat\omega}{2A} \pm k-ik\right) N\nonumber \\+\left\{ \left( (2k+1)- \frac{i\mp 1}{A}\, \hat\omega \right) y - \frac{i \pm 1}{A}\, \hat\omega \right\} N' =0,
\end{eqnarray}

which is the Hypergeometric equation with the solution

\begin{equation}\label{2f1}
 N(y)={}_2F_1 ( a, b; c; (y+1)/2),
\end{equation}

where

\begin{equation}\label{hypergeom_param}
 a=k-\frac{i \mp 1}{2A} \hat \omega+k,~b=-\frac{i \mp 1}{2A}\hat \omega,~ c=\frac{2k+1}{2} \pm \frac{\hat \omega}{A}.
\end{equation}

In order to match the behavior of the solution (\ref{Psi_ansatz_pm1}) at infinity with that demanded by 
(\ref{boundary_condition}), we demand that $N(y)$ be a polynomial as $y \to \infty$. That condition is satisfied when

\begin{equation}\label{condition}
 a=-n,~~~n=1,2,...
\end{equation}

If $a=-n$, then, according to the property of the hypergeometric equation, $N(y) \sim y^n = y^{-a}$, so that, 
according to (\ref{Psi_ansatz_pm1}),

\begin{eqnarray}
 \Psi \sim (y-1)^{-i\hat\omega/2A} (y+1)^{\pm\hat\omega/2A} y^{-a} \nonumber \\
 \approx y^{-i\hat\omega/2A} y^{\pm\hat\omega/2A} y^{-a}=y^{-2k}, 
\end{eqnarray}

as required. We deduce the expression for the asymptotic form of quasi normal frequencies from (\ref{condition}) 
as follows:

\begin{eqnarray}
 a=-n\Rightarrow 2k-\frac{i \mp 1}{2A} \hat \omega =-n \nonumber \\
 \Rightarrow \hat \omega_{asy} = A(n+2k)(\pm 1 -i) \label{asymp_qnf_hat},
 \end{eqnarray}

so that (\ref{dimless_param}) implies

\begin{equation}\label{asymp_qnf}
 \omega_{asy} = A\left(\frac{r_h}{R^2}\right)(n+2k)(\pm 1 -i),
\end{equation}

which gives the asymptotic form of the quasi normal frequencies for large maximally symmetric AdS black holes 
in the Lovelock model. We observe that the high-overtone quasi normal frequencies are equispaced, which is in agreement with 
earlier observations \cite{hubeny,berti,konoplya1}.

\subsection{Area Spectrum}\label{area_spectrum}

We calculate the area spectrum of the large AdS black holes in Lovelock model using the new physical interpretation 
of the quasi normal modes proposed
by Maggiore \cite{maggiore} and following Kunstatter's method \cite{kunstatter}. According to the first law of black
 hole thermodynamics, for a black hole system with energy $E$ (or $M$) with Hawking temperature $T_H$ and horizon area 
$A$, the following relation holds:

\begin{equation}\label{first_law}
 dM=\frac{1}{4}T_HdA.
\end{equation}

According to Kunstatter, if the frequency of oscillation of the system is $\omega (E)$, then the quantity

\begin{equation}\label{adiab_inv_E}
 I=\int \frac{dE}{\omega(E)},
\end{equation}

is to be taken as the corresponding adiabatic invariant. According to Maggiore \cite{maggiore}, the black hole system 
has to be modeled by a collection of damped harmonic oscillators. If the system has a quasi normal frequency 
$\omega = \omega_R + i \omega_I$, then the corresponding vibrational frequency according to the model is to be taken as 

\begin{equation}\label{omega0}
 \omega_0 = \sqrt{\omega_R^2+\omega_I^2}.
\end{equation}

In the highly damped and highly excited cases, $\omega_0$ can be approximated by $\omega_I$ and $\omega_R$ respectively. 
Here, we see that for higher values of the number $n$, the both $\omega_R$ and $\omega_I$ increase. Therefore we 
consider transitions between two adjacent energy levels of the system and take the physical frequency as equal to 
the difference in $\omega_0$ for the two systems; that is, we take 

\begin{equation}\label{delta_omega}
 \omega(E) = \Delta \omega = (\omega_0)_n - (\omega_0)_{n-1}.
\end{equation}

We deduce the area spectrum from (\ref{adiab_inv_E}) using the relations (\ref{horizon}) and (\ref{asymp_qnf}). The 
expression for the adiabatic invariant now reads

\begin{equation}\label{new_adiab_inv}
 I=\int \frac{dE}{\omega(E)} = \int \frac{dM}{\Delta \omega} = \int \left(\frac{dM}{dr_h}\right)\left(\frac{1}{\Delta \omega}\right)dr_h.
\end{equation}

Using (\ref{horizon}), (\ref{asymp_qnf}), (\ref{omega0}) and (\ref{delta_omega}) in (\ref{new_adiab_inv}), it is 
easy to see that the Bohr-Sommerfeld quantization condition, namely $I= n \hbar$ now reads

\begin{equation}\label{adiab_inv_exp}
 \left( \frac{d-1}{d-2}\right)\left(\frac{1}{2 \sqrt 2 A G_k R^{2k-2}}\right)r_h^{d-2} = n \hbar,
\end{equation}

which, in terms of the area $\mathcal{A}$ of the horizon, can be written in the form

\begin{equation}\label{area_spec}
 \mathcal{A} = \gamma n \hbar,
\end{equation}
where

\begin{equation}
\gamma = \Biggl[ \Gamma \left( \frac{d-1}{2} \right) \left( \frac{d-1}{d-2} \right) \left( \frac{1}{4 \sqrt 2 \pi ^{(d-1)/2} A G_k R^{2k-2}}\right)\Biggr]^{-1}.
\end{equation}

Thus the area spectrum, and consequently the entropy spectrum, of large black holes in asymptotically AdS Lovelock 
spacetimes is seen to depend on the parameters $G_k$ and the AdS radius $R$ of the theory. The dependence of $\mathcal{A}$
on $R$ is observed only when higher orders are considered. $\mathcal{A}$ is also dependent on the dimension $d$ of the spacetime, and consequently on the order 
of the theory.

\section{Conclusion}\label{conclusion}

We have analyzed the evolution of massless Klein-Gordon field in the maximally symmetric asymptotically AdS spacetime 
surrounding a black hole in the Lovelock model. We have used the form of the metric that has been derived in \cite{bhscan} in order to compute 
the quasi normal frequencies using the Horowitz-Hubeny method. The results of the numerical computation show that the
 modes in the case of higher order theories are purely damped. For the case of large as well as intermediate black 
holes, the frequencies are observed to scale linearly with the temperature of the event horizon. When we consider 
higher order theories, the imaginary part of the quasi normal frequencies is observed to be independent of the order 
of the theory in higher dimensions. They appear to be dependent on the dimension only.

The asymptotic form of the quasi normal frequencies for the case of very large black holes has been analytically 
determined using the method of perturbative expansion of the wave equation in terms of $\omega/T_H$, as developed 
in \cite{musiri}. We find that the asymptotic modes are equispaced, in agreement with previous results. We have 
also calculated the area spectrum spacing and found it to be dependent on the value of $R$, $d$ 
and $k$. This is also in contrast to the case of first order theories where we always obtain area spectra that are 
equidistant even when the parameters of the black hole spacetime change.

\section{Acknowledgement}\label{ack}

The authors would like to acknowledge financial assistance from the Council of Scientific and Industrial Research 
(CSIR), India during the work under the CSIR Emeritus Scientistship Scheme sanctioned to one of the authors (VCK). 
VCK would also like to acknowledge Associateship of IUCAA, Pune, India.

\end{document}